\documentclass[twocolumn,aps,english,prb,showpacs,longbibliography,superscriptaddress]{revtex4-1}
\usepackage{natbib}
\usepackage{float}
\usepackage[dvips,final]{graphicx}
\usepackage{color}
\usepackage{amsmath}
\usepackage{mathrsfs} 
\usepackage{graphicx, subfigure}
\usepackage{pslatex}
\usepackage{relsize}
\usepackage{bm}
\usepackage{xspace} 
\usepackage{epsfig,graphicx,amsfonts,amsbsy}
\usepackage{amsmath,amsfonts,amsthm,amssymb,mathrsfs,bbm}
\usepackage[colorlinks=true,linkcolor=blue,citecolor=blue]{hyperref} 
\usepackage{extramarks}
\usepackage{xcolor}
\usepackage{soul}

\bibpunct{[}{]}{,}{n}{}{} 

\begin{document}
\title{Oscillatory behavior of the domain wall dynamics in a curved cylindrical magnetic nanowire}


\author{R. Moreno}
\affiliation{Instituto de Ciencia de Materiales de Madrid, CSIC, Cantoblanco, \mbox{28049 Madrid, Spain}}
\affiliation{Department of Physics, University of York, Heslington, York YO10 5DD, United Kingdom}
\author{V. L. Carvalho-Santos}
\affiliation{Instituto Federal de Educa\c c\~ao, Ci\^encia e Tecnologia Baiano, {48970-000, Senhor do Bonfim, Brazil}}
\author{A. P. Espejo}
\affiliation{Departamento de F\'{i}sica, CEDENNA, Universidad de Santiago de Chile, Av. Ecuador $3493$, Santiago, Chile.}
\author{D. Laroze}
\affiliation{Instituto de Alta Investigaci\'{o}n, CEDENNA, Universidad de. Tarapac\'{a}, Casilla 7D, Arica, Chile.}
 \affiliation{School of Physical Sciences and Nanotechnology, Yachay Tech University, 00119-Urcuqu\'{i}, Ecuador.}
\author{O. Chubykalo-Fesenko}
\affiliation{Instituto de Ciencia de Materiales de Madrid, CSIC, Cantoblanco, \mbox{28049 Madrid, Spain}}
\author{D. Altbir}
\affiliation{Departamento de F\'{i}sica, CEDENNA, Universidad de Santiago de Chile, Av. Ecuador $3493$, Santiago, Chile.}
\email[]{dora.altbir@usach.cl}


\begin{abstract}

Understanding the domain wall dynamics is an important issue in modern magnetism. Here we present results of domain wall displacement in curved cylindrical  nanowires at a constant magnetic field. We show that the average velocity of a transverse domain wall increases with curvature. Contrary to what it is observed in stripes, in a curved wire the transverse domain wall oscillates along and rotates around the nanowire  with the same frequency. These results open the possibility of new oscillation-based applications.
    
\end{abstract}
\pagebreak
\maketitle


\section{Introduction}
\label{S1}

 The understanding of the domain wall (DW) dynamics is a cornerstone in nanomagnetism owing to the potential technological applications in modern nanoelectronic such as high density memories \cite{Parkin} and logic devices \cite{Allwood}. In particular, the promising concept of ``race-track memory'' \cite{Parkin,Catalan} demands a well controlled motion of the DW along a nanowire. In this context, several works have  studied  the controlled displacement of DWs in stripes \cite{Beach-Nat-2005,Miron-Nat-2011,Emori-Nat-2013,Kim-Nat-2014,Porter,Mougin} and cylindrical  wires \cite{Hertel2,Hertel1,Wieser}. Such studies have revealed that geometry plays  a fundamental role in the DW dynamics. For instance, rectangular wires exhibit a Walker field \cite{Walker}, that is, while at low fields the DW velocity is linearly proportional to the field, at some critical value the velocity drastically drops  and an oscillatory behavior of the DW position is observed \cite{Porter}. { On th
 e contrary}, in cylindrical nanowires, the DW does not change its structure during the motion and thus no Walker limit or critical velocity is observed  \cite{Hertel1}. Other examples of important geometrical effects on the DW motion are an increase of the DW velocity by a factor of four if a magnetic stripe has a series of cross shaped traps \cite{Lewis-NMat-2010} or the pinning of DW by artificial necks in wires \cite{Himeno-JAP-2003}.

Since the shape of the nanowire, frequently  curved by nature, affects the magnetization statics and dynamics \cite{Gaididei-PRL-2014}, and as well as many of potential application of DW include curvilinear segments along the wire (see for Example Ref. \cite{Parkin}), it is fundamental to understand how the curvature influences the DW dynamics. Recent theoretical studies have shown that new interesting phenomena appear when a cylindrical wire is curved. Firstly, it was shown that due to a curvature-induced Dzyaloshinsky-Moriya interaction (DMI), the DW gets pinned at the maximum of the curvature and, for minimizing exchange cost, a phase selectivity occurs, that is, a head-to-head DW is directed outward while a tail-to-tail is directed inward the bend  \cite{Yershov}. This curvature-induced DMI is also responsible for the formation of a new DW profile, given by a head-to-head vortex-antivortex pair in toroidal magnetic nanoparticles \cite{Vagson-JAP-Submitted}. {Recently,
  the effects of the curvature on a magnetic helicoid ribbon as well as on a M\"obius ribbon have been analyzed \cite{Gaididei17}}. It was also shown that the spin-current driven DW motion is strongly dependent on curvature and torsion \cite{Yershov2}. In this case, the curvature results in the existence of a Walker limit for uniaxial wire, and the torsion induces an effective shift of the nonadiabatic spin torque parameter \cite{Yershov2}. The direction of motion of a DW along a helical wire under the action of a Rashba spin orbit torque depends on the helix chirality and wall charge in such way that DWs can be moved only under the action of Rashba and geometrical effects \cite{Pylypovskyi}. Thus, the curvature can induce inhomogeneities in the DW profile. Under this frame, the study of the influence of curvature on the DW velocity is a paramount. 

In this paper, by means of an analytical model and micromagnetic simulations, we explore the dynamics of a transverse DW in a curved cylindrical nanowire driven by a constant external magnetic field. We have analyzed the DW motion for different curvatures of the nanowire, going from 0 until it reaches its maximum, that is a half torus section. Our results demonstrate a strong dependence of the DW dynamics on the curvature of the wire leading to the possibility of engineering its characteristic features. The manuscript is organized as follows: In Sec. \ref{S2}, the analytical model for a curved nanowire is developed and the effective equation of motion for the DW are presented. In Sec. \ref{S3}, numerical simulations using NMAG micro-magnetic software \cite{NMAG} are performed for different control parameters. Finally, conclusions are presented in Sec. \ref{S4}.

\section{Theoretical model} 
\label{S2}

Let us consider a curved wire with length $L$ and diameter $d$ characterized by a magnetization, $\mathbf{M}$. Its dynamics is modeled by the Landau-Lifshitz-Gilbert equation
\begin{equation}
\label{LLGEq}
\frac{\partial \mathbf{m}}{\partial \tau}=\mathbf{m}\times\frac{\delta\mathcal{E}}{\delta\mathbf{m}} + \alpha\,\mathbf{m}\times\frac{\partial\mathbf{m}}{\partial \tau},
\end{equation}
where $\mathbf{m}=\mathbf{M}/M_S$, such that  $M_{S}$ is the saturation magnetization. Here, we use dimensionless time $\tau= t\gamma_0| M_s$ and dimensionless energy $\mathcal{E}=E/(\mu_0M_S^2 V)$, such that $V$ is the volume, $\mu_0$ is the magnetic permeability, $\gamma_0=\mu_0|\gamma|$ with $\gamma = 1.76\cdot 10^{11} rad/T\cdot s$ being the gyronagnetic factor. Additionally, $\alpha$ is the dimensionless Gilbert damping parameter. In our calculations, we consider a Permalloy wire characterized by a saturation magnetization  $ \mu_0 M_s = 1T$ and exchange  constant $ A =1.3\cdot 10^{-11} J/m$ and we fix the damping constant to $\alpha =0.01$. 

\begin{figure}[h!]
\centering
\includegraphics[width=0.8\columnwidth ,angle=0,trim=2 10 2 2,clip]{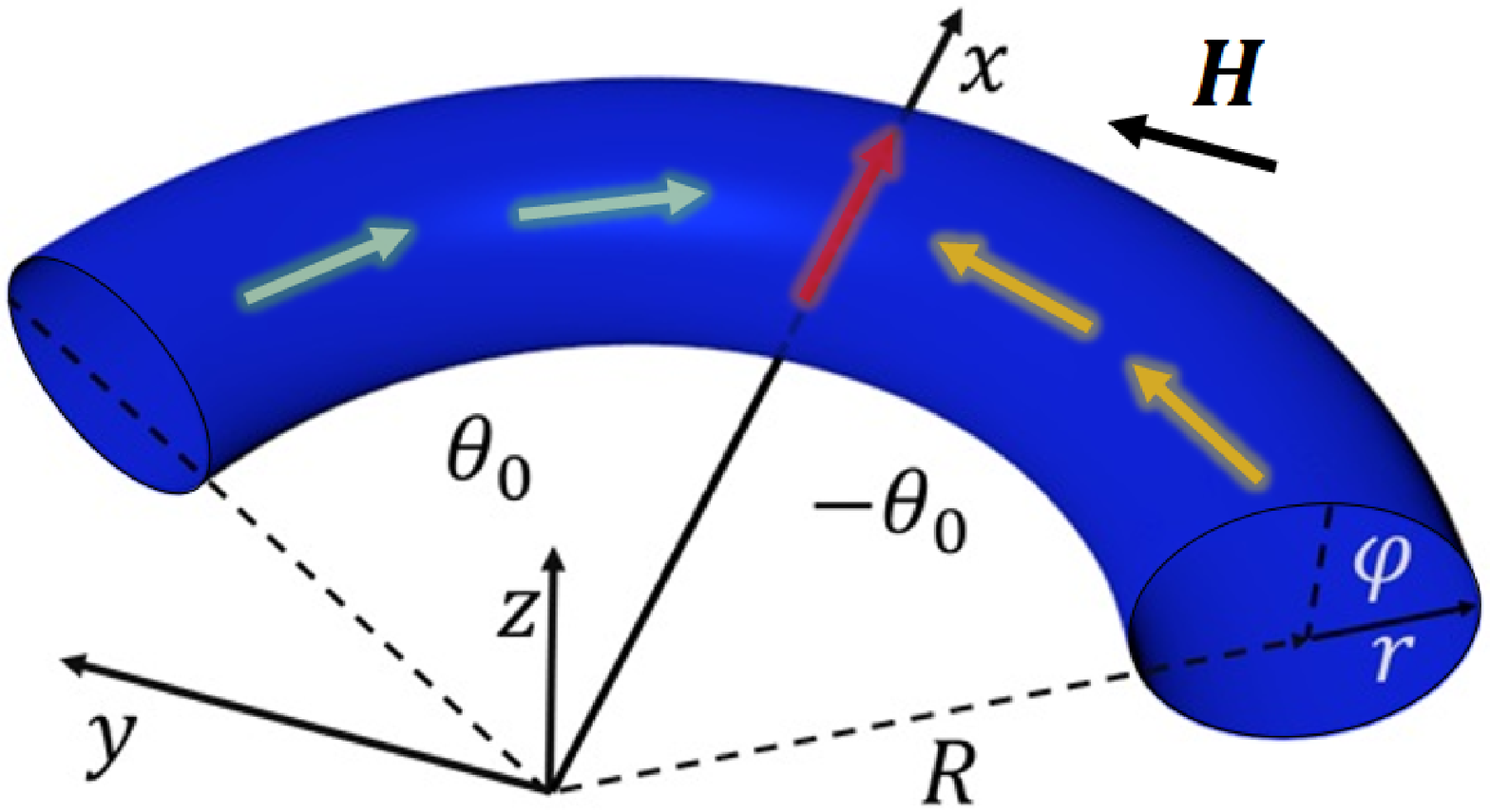}
\includegraphics[width=0.8\columnwidth ,angle=0,trim=2 10 2 2,clip]{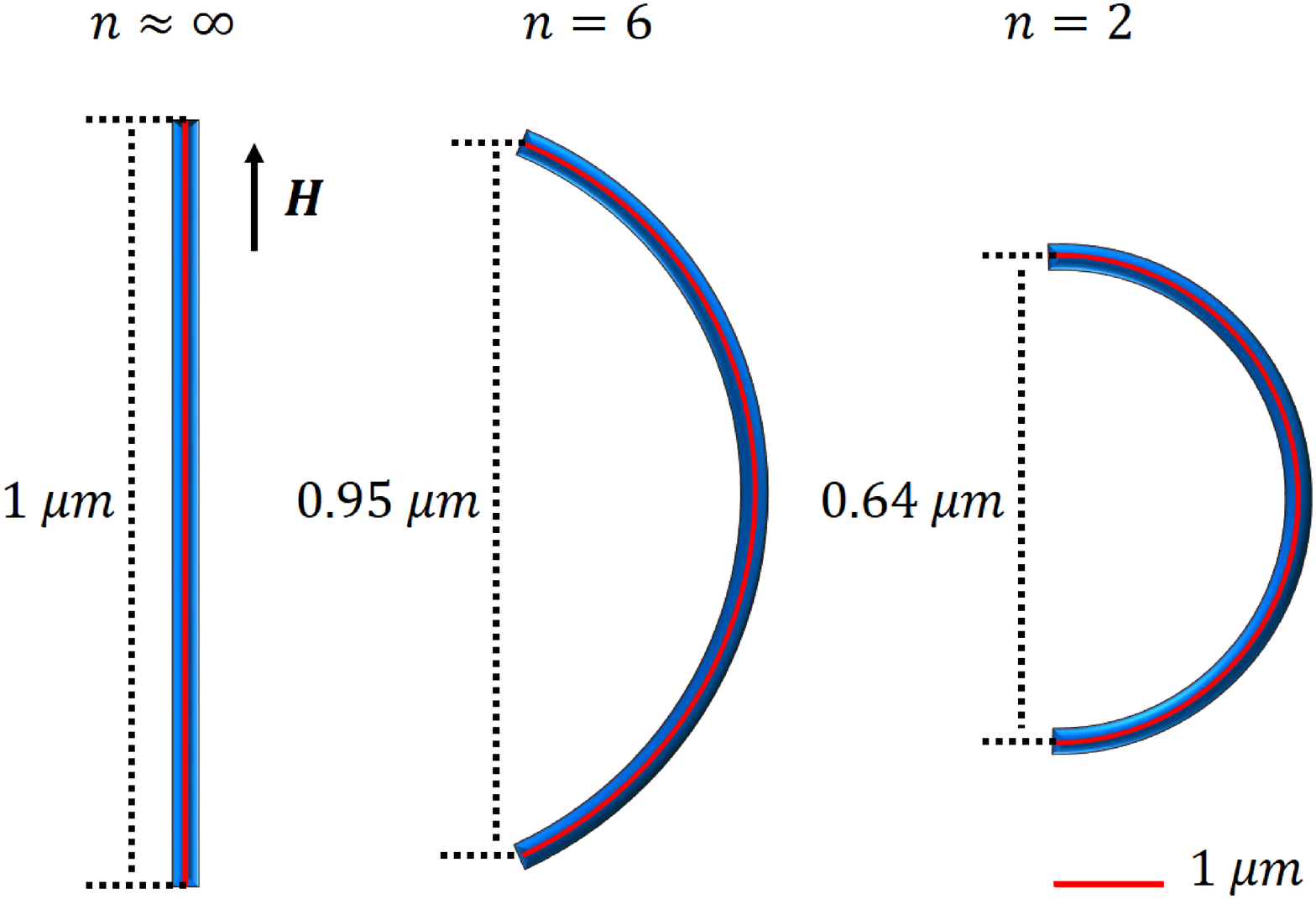}
\caption{Top figure shows the adopted coordinate system to describe the curved wire, the magnetization direction and the domain wall. Bottom figures show different nanowires with its corresponding $n$ value and dimensions. All the wires have $1 \mu m$ of length along their axis, which is indicated with red lines.}
\label{Fig1}
\end{figure}


The geometrical description of a curved nanowire is done by using a toroidal coordinate system, given by
\begin{equation}
\label{CoordSystem}
\mathbf{r} = \hat{e}_R\,(r + R \cos \theta) \, + \, \hat{\theta}\,R \sin \theta  \,, 
\end{equation}
where $\hat{e}_R$ and $\hat{\theta}$ are the radial unitary vector and the angular unitary vector, respectively, with $R$ and $r$ as the toroidal and poloidal radii, respectively. The angles $\theta\in[-\theta_0,\theta_0]$ and $\varphi\in[0,2\pi]$ play the role of azimuthal and poloidal angles, respectively, as it is shown in Fig. \ref{Fig1}. This parametrization allows us to define the wire length as $L=2\theta_0 R$, which is fixed in this work. Even though this parametrization yields a geometry with variable Gaussian curvature, in this work we will define the curvature of the wire as $K=1/R$. Then, variations in the curvature will be represented by simultaneous changes in the toroidal radius $R=nL/2\pi$ and azimuthal angle $\theta \in [-\pi/n:\pi/n]$, where $n\in [2,\infty)$ is a real number (See Fig. \ref{Fig1}). It can be noted that the wire with greater curvature, described by a half-torus section, is obtained when $\theta_0=\pi/2$, and an almost straight wire is obtained for $n \rightarrow \infty$. The description of such curved wire by its toroidal and poloidal radii is more  suitable than other representations \cite{Gradshtein-Book,Vagson-JAP-2010,Bellegia-PhySoc-2008}. 

Let us parametrize the magnetization $\mathbf{m}$ in the spherical coordinate system lying on a curvilinear background described by a Frenet-Serret basis ($\hat{\theta},-\hat{e}_R,\hat{z}$), that is, $\mathbf{m}=\hat{\theta}\cos\Theta-\hat{e}_R\sin\Theta\cos\Phi+\hat{z}\sin\Theta\sin\Phi$, where the directional vectors are associated with a toroidal coordinate system given in Eq. (\ref{CoordSystem}). 

The LLG equation in this coordinates can be written as 
\begin{equation}
\label{LLGs1}
-\sin\Theta \frac{\partial \Theta}{\partial \tau}=\frac{\delta \mathcal{E}}{\delta \Phi}+\alpha\,\sin^2 \Theta \,\frac{\partial \Phi}{\partial \tau},
\end{equation}
and
\begin{equation}
\label{LLGs2}
\sin\Theta \frac{\partial \Phi}{\partial \tau}=\frac{\delta \mathcal{E}}{\delta \Theta}+\alpha\,\frac{\partial \Theta}{\partial \tau}.
\end{equation}
For the energy, we employ a simple model that contains only tree contributions: exchange, magnetostatic and Zeeman. Hence, the energy can be written as
\begin{equation}
\label{TotMagEnergy}
{{E}}=S\,R\int_{\,-\theta_0}^{\,\theta_0}[\ell^2 {E}_{ex}-\lambda\cos^2\Theta-\frac{\ell^2}{A}\mathbf{H}\cdot\mathbf{M}]\,d\theta\,,
\end{equation}
where $S=\pi r^2$ is the area of the wire cross section, $\ell=\sqrt{A/(\mu_0 M_S^2)}$, $\mathcal{E}_{ex}$ is the exchange energy density, $\mathbf{H}=H(\hat\theta\,\cos\theta-\hat{e}_R\,\sin\theta)$ is the external magnetic field pointing along the y-axis direction,  and $\lambda$ is the dimensionless anisotropy constant. Following the model described in Refs. \cite{Yershov,Yershov2}, we will consider that $\lambda>0$ represents an easy-tangential anisotropy coming from magnetostatic contributions. It is worth noting that there is a dependence of $\lambda$ on $\theta$ and $\varphi$, coming from the fact that the demagnetizing field of a curved wire depends on the direction in which the magnetization field is pointing to \cite{Vagson-development}. However, because of the considered dimensions, this difference is small, and we will adopt $\lambda=1/4$ to describe magnetostatic contributions to the energy \cite{Porter,Yershov,Yershov2}. In terms of the adopted parametrization, the exchange energy density is given by \cite{Sheka-JPA-2015,Yershov} 
\begin{equation}
\label{EexDensity}
\frac{{E}_{ex}}{K^2} = \left(\frac{\partial \Theta}{\partial \theta}+\cos \Phi \right)^2+\left(\sin \Theta \frac{\partial \Phi}{\partial \theta} - \cos \Theta \sin \Phi \right)^2 .
\end{equation}
The DW properties can be analyzed based on a collective variable approach \cite{Slonk,Thiele}. In this manner, a {\it head-to-head} DW can be described by the ansatz
\begin{equation}
\label{DWSolution}
\Theta(\theta,\tau)=2\arctan \left[\exp{\left(\frac{\zeta(\theta,\tau)}{\Delta}\right)}\right]\,,\,\,\,\,\,\,\,\Phi(\tau)=\phi(\tau),
\end{equation}
where $\zeta(\theta,\tau)=R\theta-q(\tau)$. Here $q$ and $\phi$ denote the canonically conjugated pairs of collective variables, determining the DW position and phase, respectively, and  $\Delta$ determines the DW width, which is assumed to be constant and independent on the DW position or the direction. This approximation is valid once the considered geometry does not present torsion and a generalized DW model \cite{Kravchuk-JMMM-2014} is not necessary. From Eqs. (\ref{EexDensity}) and (\ref{DWSolution}), it can be noted that in the absence of an external magnetic field, the minimum energy density is obtained for $\Phi_0=\pi$ and the DW has its minimum energy when it points along $\hat{e}_R$ (outward the bend), according to Yershov {\it et al.} \cite{Yershov}. 

\begin{figure}[h]
\includegraphics[width=0.8\columnwidth ,angle=0]{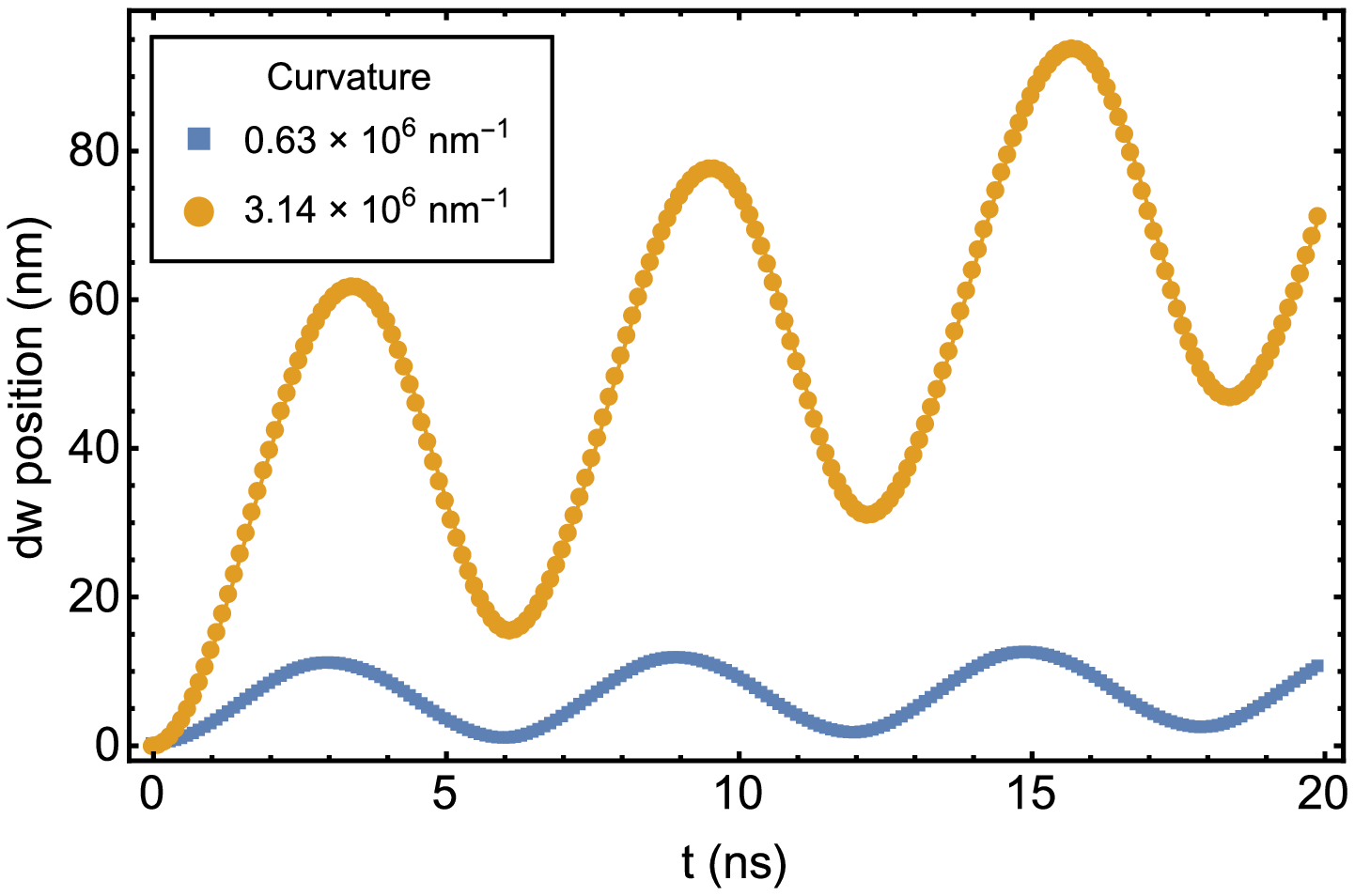}
\includegraphics[width=0.8\columnwidth ,angle=0]{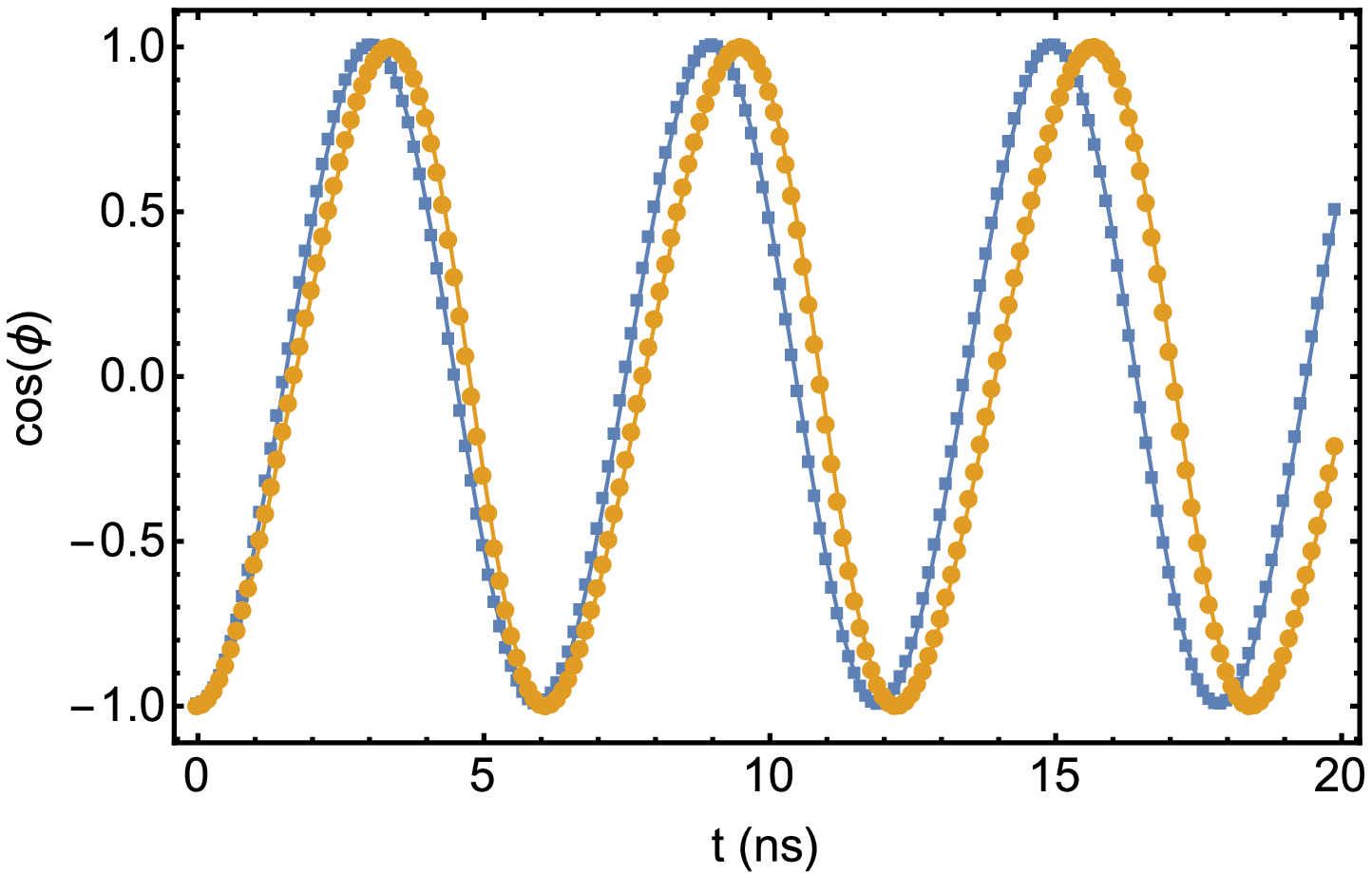}
\caption{Domain wall  position, $q$ (top), and phase, $\phi$ (bottom), obtained from the numerical integration of equations (\ref{DwDynamics1}) and (\ref{DwDynamics2}). { The used value of the magnetic field is $H= 6$ mT.}}
\label{Fig2}
\end{figure}

The energy of the DW can be obtained analytically by performing the substitution of Eq. (\ref{DWSolution}) in (\ref{TotMagEnergy}) but its expression is cumbersome { and therefore is presented in the Appendix}. The dynamical properties of the domain wall in terms of the collective variables can be described by the following equations of motion \cite{Yershov}
\begin{equation}
\label{DwDynamics1}
\dot{q}=\frac{1}{2S}\frac{\partial\mathcal{E}_{DW}}{\partial\phi}+\alpha\,\Delta\,\dot{\phi}\,,
\end{equation}
and
\begin{equation}
\label{DwDynamics2}
\dot{\phi}=-\frac{1}{2S}\frac{\partial\mathcal{E}_{DW}}{\partial q}-\frac{\alpha}{\Delta}\,\dot{q}\,,
\end{equation}
which have been derived inserting the ansatz (\ref{DWSolution}) in Eqs. (\ref{LLGs1}) and (\ref{LLGs2}), respectively. The above equations do not have simple analytical solutions; nevertheless, they can be solved numerically. The obtained  DW position and phase are shown in Fig. \ref{Fig2} { for a fixed value of the magnetic field $H= 6$ mT}.  The upper frame shows the oscillatory behavior of the DW center along the wire. The lower frame presents the DW phase for the angular velocity of the DW rotation around the wire. The period of the DW displacement is approximately 6 ns, and in this time interval the phase changes in $2\pi$, giving a precessional frequency of $60^0$ per nanosecond. This behavior of the DW direction is different to stripes that present a $180^0$ rotation around the nanowire followed by a inversion of the sense of the rotation.  This difference can be explained because nanostrips have two minima associated with the magnetostatic energy when the DW is pointing along the thinner sides of the stripe (exchange energy is equal during all the DW motion). On the other hand, the DW magnetic energy presents only one minimum in a curved cylindrical wire, which is mainly associated with the exchange energy. However, in curved cylindrical wires the DW has different values of the exchange energy when it is pointing in or outward the bent of the wire. Then, the DW needs to rotate an angle $2\pi$ before recovering its minimum energy position. {Let us also remark that unlike its straight counterpart, curved wires present the oscillatory behavior associated with the Walker Breakdown phenomenon. As explained in Ref. \cite{Hertel1}, the Walker breakdown
is absent in straight cylindrical wires because the demagnetizing factors and the exchange effective field do not depend on the DW phase. Therefore the contribution coming from the magnetostatic and exchange energies to the effective fields acting on the domain wall (which produces the Walker breakdown) is cancelled upon derivatives and do no contribute to the DW displacement. In the case of a curved wire, the corresponding effective fields depend on the DW phase (See Appendix) and therefore,  there will be always an  oscillatory behavior in the DW position during its displacement.}

\begin{figure}[h!]
\centering
\includegraphics[width=1.0\columnwidth ,angle=0]{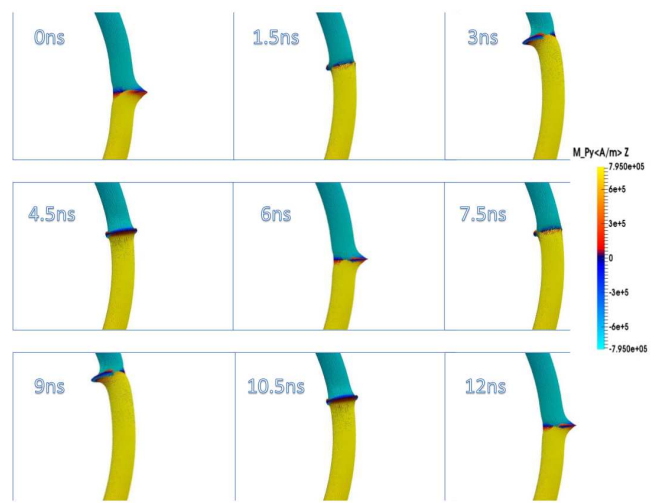}
\caption{Schematic representation for the oscillatory movement of the DW pushed by a constant magnetic field. The wire has its maximum curvature $1/R\approx 3.14 \cdot 10^6 nm^{-1}$.}
\label{Fig3}
\end{figure}

We remark that our model provides quantitative results which are valid not far from the nanowire center. Since the magnetic field is applied along the y-axis direction, the torque produced by the field is maximum when the domain wall is at the center of the magnetic wire. When we are far from the center, the domain wall experiences a smaller torque and, although the qualitative behavior is exactly the same, the velocity slows down. 

\section{Numerical Results}  
\label{S3}

To better understand our analytical results, we developed several full numerical simulations. We use the public micromagnetic code NMAG \cite{NMAG}, which considers finite element discretization, allowing a better description of curved surfaces compared to finite differences methods. With this code we solve the Landau-Lifshitz-Gilbert equation (\ref{LLGEq}) for a Py nanowire. Since we  focus on a {\it head-to-head} DW, position of the center of the DW is obtained from $m_{\|}(\textbf{r}_{DW},t)=0$, where $m_{\|}$ is the parallel component along the  wire axis.  Using the angular difference $\bigtriangleup\theta$ between the position of the center of the DWs  at two consecutive times, we define the instantaneous velocity as $\textrm{V}(t)= R\cdot\bigtriangleup\theta/\bigtriangleup t$. All the simulated wires have same diameter ($d=30\,$nm) and length ($L =1\,\mu$m), but different curvatures ($K$),  which were varied as a function of $n$. The initial state of all simulations is a trans
 verse DW in a wire, obtained by saturating the system in the +x direction with a $1T$ field. Then, the field is switched off, and the system is  left to relax to its equilibrium state, as explained in Ref. \cite{Saitoh}. In this way we create a  head-to-head DW pointing along the  +x direction (outward the bend), as predicted in Ref. \cite{Yershov}. Starting from this state, we apply a constant magnetic field along the +y direction and perform the simulation for $20$ ns. { To check our methodology we have obtained the velocity from our simulations in straight wires and compared our results with the velocity obtained in the literature \cite{Hertel1} for similar systems. Both results are in good agreement, giving confidence in the software and method used in our calculations}.

Fig. \ref{Fig3} illustrates  snapshots at different times of the reversal process of the DW propagation along a wire with curvature $K\approx 3.14 \cdot 10^6 nm^{-1}$. In these snapshots a  rotation of the DW around the wire axis as well as an oscillation along its axis is shown. It should be highlighted that  the sense of the rotation of the DW is constant. The DW  oscillations along the propagation line have been observed in previous works for DW dynamics in nanostripes under the action of a constant magnetic field \cite{Porter,Mougin,Guslienko}. However, in the case of a Py stripe, the DW displacement oscillates with double frequency than the rotation frequency of the wall around the stripe \cite{Porter,Mougin}. In the present study, as is clearly seen in Fig. \ref{Fig3}, and in agreement with our theoretical predictions, we have observed that the period of the DW displacement along the wire is exactly the same as the rotation period around it. 

\begin{figure}[h!]
\includegraphics[width=1.0\columnwidth ,angle=0]{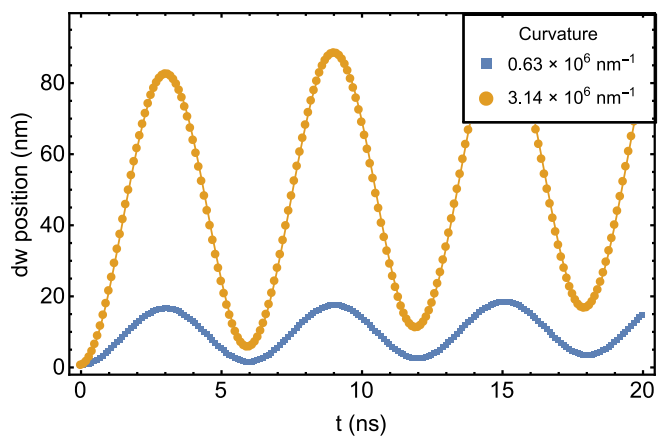}
\caption{Displacement of the domain wall center as a function of time for two different curvatures at $H= 6$ mT found using NMAG simulations.}
\label{Fig4}
\end{figure}

The dynamics of the DW position  is shown in Fig. \ref{Fig4}. We observe that it has an oscillatory behavior and its amplitude depends on the curvature also in agreement with the analytical model.  In fact, both approaches exhibit the same oscillation period. The discrepancy in the exact DW position obtained from both methods  comes from the fact that the theoretical model fixes the width of the DW, however  simulations show a small change in the width during rotation.

\begin{figure}[h!]
\centering
\includegraphics[width=0.9\columnwidth ,angle=0]{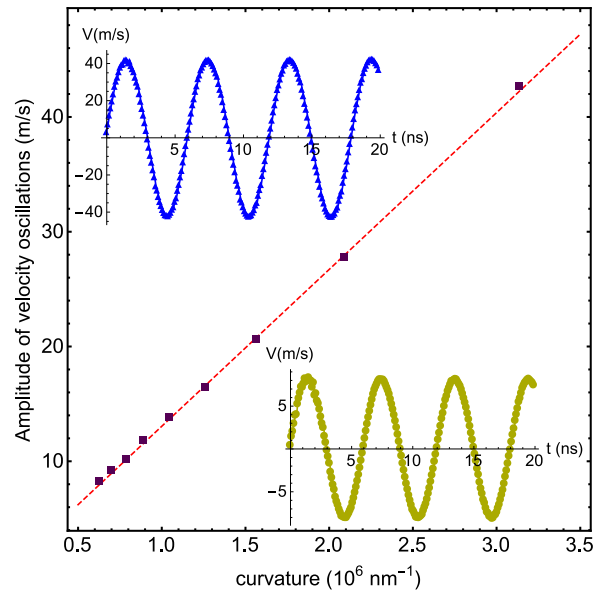}
\caption{{ Ampitude of velocity oscillation} as a function of the curvature at $H= 6$ mT. {Insets show the velocity of the DW for wires with curvatures $3.14 \times10^6$ nm$^{-1}$ (Blue triangles) and $0.63 \times10^6$} nm$^{-1}$ (Beige dots)}.
\label{Fig5}
\end{figure}
Despite the existence of an oscillatory behavior on the DW position, an increase in the curvature implies an increment in average { ampitude of velocity oscillation}. This occurs because, as shown in Fig. \ref{Fig1}, the change in the direction of a magnetic moment when passing the DW is smaller in a curved wire, as compared to what occurs in a straight structure. Based on our results, we have obtained the amplitude of the DW velocity as a function of the wire curvature. Our simulations show that the { ampitude of velocity oscillation} does not depend on the applied magnetic field, varying only with the shape of the nanowire. In fact, the amplitude of the velocity is a linear function of the curvature, as shown in Fig. \ref{Fig5}.

\begin{figure}[h!]
\includegraphics[width=0.9\columnwidth ,angle=0]{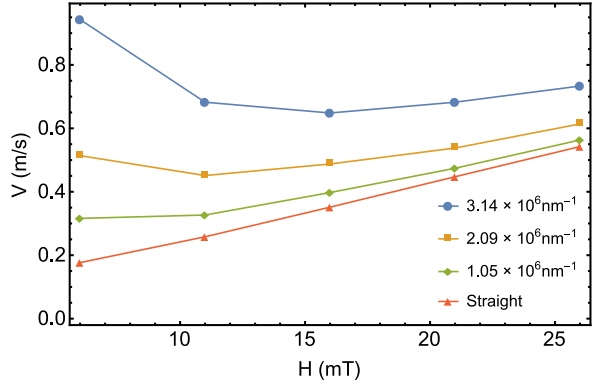}
\caption{Average velocity as a function of the external magnetic field for different curvatures.}.
\label{Fig6}
\end{figure}

Finally, Fig. \ref{Fig6} shows the average of the DW velocity as a function of the external field for different  curvature values. We can observe that the average velocity is no longer a linear function of the magnetic field when the wire is curved.  Increasing the curvature produces a more noticeable change of the linear behavior. Our modeling show that at very small fields of the order of  $1$ mT (not shown due to their too small values) the linear regime is recovered in these nanowires. In this sense, the Walker breakdown, absent in straight cylindrical wires \cite{Hertel1}, is observed in nanowires with large curvature.

\section{Conclusions} 
\label{S4}

 In this paper we studied domain wall propagation in curved cylindrical nanowires showing an oscillatory behavior along and around the nanowire axis. Contrary to the case of rectangular stripes, the rotational period was proven to be the same as the oscillations in the DW position. The amplitude of the velocity depends on the magnitude of the demagnetizing field that, in turn, depends on the curvature. The frequency of these oscillations is a linear function of  the external magnetic field. The average velocity is not linear in the applied field and, contrary to what happens in stripes, curved nanowires show the Walker breakdown phenomenon. The oscillation of the domain wall around and along the nanowire provide a unique opportunity to convert constant magnetic field to radio-frequency signal with tunable parameters coming from nanowire geometry, material and curvature, such oscillations should be also possible in the presence of spin-currents which could open novel possibilities for
  spin-torque oscillators based on magnetic nanowires.

\section*{Acknowledgements}

In Chile, we thank the par- tial financial support from Centers of excellence with BASAL/CONICYT financing, Grant FB0807, Fondecyt (1160198) and CONICYT-ANILLO ACT 1410. In Brazil, we thank agencies CNPq (grant No. 301015/2015-5) and Fapesb (grant No. JCB0063/2016). Spanish Ministry of Economy and Competitiveness (MAT2013-47078-C2-2-P, MAT2016-76824-C3-1-R and FIS2016-78591-C3-3-R). We are also grateful to V.P. Kravchuk for fruitful discussions on the theoretical model. 

\appendix
\section{Domain wall energy}

We have considered that the magnetic energy of the wire under an external magnetic field has three contributions, $E=E_{\text{ex}}+E_{\text{dip}}+E_{\text{Z}}$, where $E_{\text{ex}}$ is the exchange energy,  $E_{\text{dip}}$ is an effective dipolar energy represented by an in-surface anisotropy, and $E_{\text{Z}}$ is the Zeeman energy. The general expressions of the energy are given by Eqs. (\ref{TotMagEnergy}) and (\ref{EexDensity}). Inserting the DW's ansatz (\ref{DWSolution}) into the aforementioned equations we obtain {

%
\begin{equation}
\label{AppExchange}
E_{\text{ex}}=\frac{\ell^2S}{R^2}\left[L+4R\,\Lambda(q)\,\cos\phi+\frac{2R^2-\Delta^2+\Delta^2\cos2\phi}{\Delta}\,\Omega(q)\right]\,,
\end{equation}
%
where 
\begin{equation}
\Lambda(q)=\arctan\left[\exp\left(\frac{-q+L/2}{\Delta}\right)\right]-\text{arccot}\left[\exp\left(\frac{q+L/2}{\Delta}\right)\right] \nonumber
\end{equation}
and 
\begin{equation}
\Omega(q)=\frac{\sinh\left(\frac{L}{\Delta}\right)}{\left[\cosh\left(\frac{2q}{\Delta}\right)+\cosh\left(\frac{L}{\Delta}\right)\right]},  \nonumber 
\end{equation}
The dipolar energy is given by an in-surface anisotropy, being evaluated as}

\begin{equation}\label{AppDipolar}
E_{\text{dip}}=2S\lambda\left[{\Delta\,\Omega(q)}-\frac{L}{2}\right]\,.
\end{equation}

Finally, the energy from the interaction with the external magnetic field is given by

\begin{widetext}
\begin{equation}\label{AppZeeman}
E_{\text{Z}}=\frac{\text{ie}^{-\text{i}\theta_0}HSR}{2}\left\{2-2\text{e}^{2\text{i}\theta_0}+(\cos\phi-1)\left[_2F_1\left(1,-\frac{\text{i}\Delta}{R},1-\frac{\text{i}\Delta}{R},-\text{ie}^{\frac{-q+L/2}{\Delta}}\right)+_2F_1\left(1,\frac{\text{i}\Delta}{R},1+\frac{\text{i}\Delta}{R},\text{ie}^{\frac{-q-L/2}{\Delta}}\right)\right]\right.\nonumber\end{equation}
\begin{equation}
-\text{e}^{2\text{i}\theta_0}(\cos\phi-1)\left[_2F_1\left(1,-\frac{\text{i}\Delta}{R},1-\frac{\text{i}\Delta}{R},-\text{ie}^{\frac{-q-L/2}{\Delta}}\right)+_2F_1\left(1,\frac{\text{i}\Delta}{R},1+\frac{\text{i}\Delta}{R},\text{ie}^{\frac{-q+L/2}{\Delta}}\right)\right]\nonumber
\end{equation}
\begin{equation}
-(\cos\phi+1)\left[_2F_1\left(1,-\frac{\text{i}\Delta}{R},1-\frac{\text{i}\Delta}{R},\text{ie}^{\frac{-q+L/2}{\Delta}}\right)+_2F_1\left(1,\frac{\text{i}\Delta}{R},1+\frac{\text{i}\Delta}{R},-\text{ie}^{\frac{-q-L/2}{\Delta}}\right)\right]\nonumber
\end{equation}
\begin{equation}\left.
+\text{e}^{2\text{i}\theta_0}(\cos\phi+1)\left[_2F_1\left(1,-\frac{\text{i}\Delta}{R},1-\frac{\text{i}\Delta}{R},\text{ie}^{\frac{-q-L/2}{\Delta}}\right)+_2F_1\left(1,\frac{\text{i}\Delta}{R},1+\frac{\text{i}\Delta}{R},-\text{ie}^{\frac{-q+L/2}{\Delta}}\right)\right]\right\}\,,
\end{equation}
\end{widetext}
where $_2F_1(a,b,c,z)$ is the Hypergeoetric function.

{Aiming to understand the difference between the DW dynamics in curved and straight wires, we will study an approximation in which the DW is near the center of the wire, and its width is much smaller than the wire length. In this case $q$ is of the order of tens nanometers, and therefore $\Delta/L\sim10^{-3}$. Consequently, Eqs. (\ref{AppExchange}) and (\ref{AppDipolar}) are simplified to
\begin{equation}
E_{\text{ex}_{\Delta/L\rightarrow0}}\approx\frac{\ell^2S}{R^2}\left[L+\frac{2R^2}{\Delta}+2\pi\,R\,\cos\phi\right]\,,
\end{equation}
\begin{equation}
\label{ApproxDip}
E_{\text{dip}_{\Delta/L\rightarrow0}}\approx-\lambda\,V\,,
\end{equation}
where $V$ is the wire volume. By using of the considered limits, the functions $\Lambda(q)$ and $\Omega(q)$ are reduced to $\Lambda(q)\approx\pi/2$ and $\Omega(q)\approx1$. Even with these approximations, the Zeeman energy keeps its dependence on $q$ and $\phi$, which results in the oscillatory behavior of DW dynamics in curved nanowires. On the other hand, a straight wire is obtained in the limit $R\rightarrow\infty$ and then, $E_{\text{ex}_{R\rightarrow\infty}}\approx{2\ell^2S}/\Delta\,$, while dipolar energy keeps the form presented in Eq. (\ref{ApproxDip}). In addition, from taking the derivatives of the Zeeman energy and adopting the limit $R\rightarrow\infty$, it can be shown that $\partial E_{Z_{R\rightarrow\infty}}/\partial q=-2HS$ and $\partial E_{Z_{R\rightarrow\infty}}/\partial \phi=0$. Therefore, substituting these results in Eqs. (\ref{DwDynamics1}) and (\ref{DwDynamics2}) and performing some algebraic manipulation, we obtain that the DW velocity in a straight wire is
\begin{equation}
\dot{q}=\frac{\alpha\Delta}{1+\alpha^2}\,H\,,
\end{equation}
which is the equation of motion for DW along a straight wire \cite{Porter,Mougin}. }


\end{document}